\documentclass[prb,twocolumn,aps,showpacs,fixfloats]{revtex4}
\usepackage{graphicx}
\usepackage{bm}
\usepackage{amsmath,amssymb}
\usepackage{subfigure}
\usepackage{float}
\usepackage{latexsym}
\usepackage{color}
\usepackage{enumerate}
\usepackage{pdfpages}
\usepackage{tikz}
\usepackage{hyperref}
\usepackage{graphicx}% complex graphics
\usepackage{bm}% bold math
\usepackage{amssymb} %math symbols
\usepackage{amsmath}
\usepackage{subfigure}

\usepackage{relsize}
\begin{document}

\title{Three-electron bunches in occupation of a 5-site Coulomb cluster}

\author{R. E. Putnam, Jr. and M. E. Raikh}

\affiliation{ Department of Physics and Astronomy, University of Utah, Salt Lake City, UT 84112}

\begin{abstract}
Attraction of like charges in a
localized system
%system
%of localized electrons
implies that,
upon increasing the Fermi energy,
%$\mu$,
the occupation of the system changes as, $n\rightarrow (n+2)$, while the occupation, $(n+1)$, is skipped. In this way, the attraction translates into
the bunching of electrons.
%With infinite on-site repulsion,
For a localized system of $N=4$ sites,
attraction of electrons manifests itself in skipping of $n=2$ occupation.
The origin of the attraction is rearrangement of the occupations of the surrounding sites which plays the role of a polaronic effect.
%the possibility of
%2e-attraction
%$2e$-bunching was previously demonstrated
%for a $N=4$-site cl
We consider an $N=5$-site cluster and demonstrate that,
with screened Coulomb repulsion, {\em three-electron} bunching becomes possible, i.e. the change of occupation $n=1\rightarrow n=4$  with $n=2$ and $n=3$ occupations skipped.

\end{abstract}

\maketitle

\section{Introduction}
A question whether two electrons can attract each other without lattice dynamics involved was previously posed in many papers.
%giving rise to phonons
Another formulation of
this question is whether the  formation of negative-$U$ centers,\cite{Anderson1975} which are
believed to be due to phonons, is possible as a result of
electron-electron repulsion.
An appeal of having purely electronic negative-$U$
centers is that such centers can be viewed as precursors of purely electronic superconducting state.
\begin{figure}
\includegraphics[width=90mm]{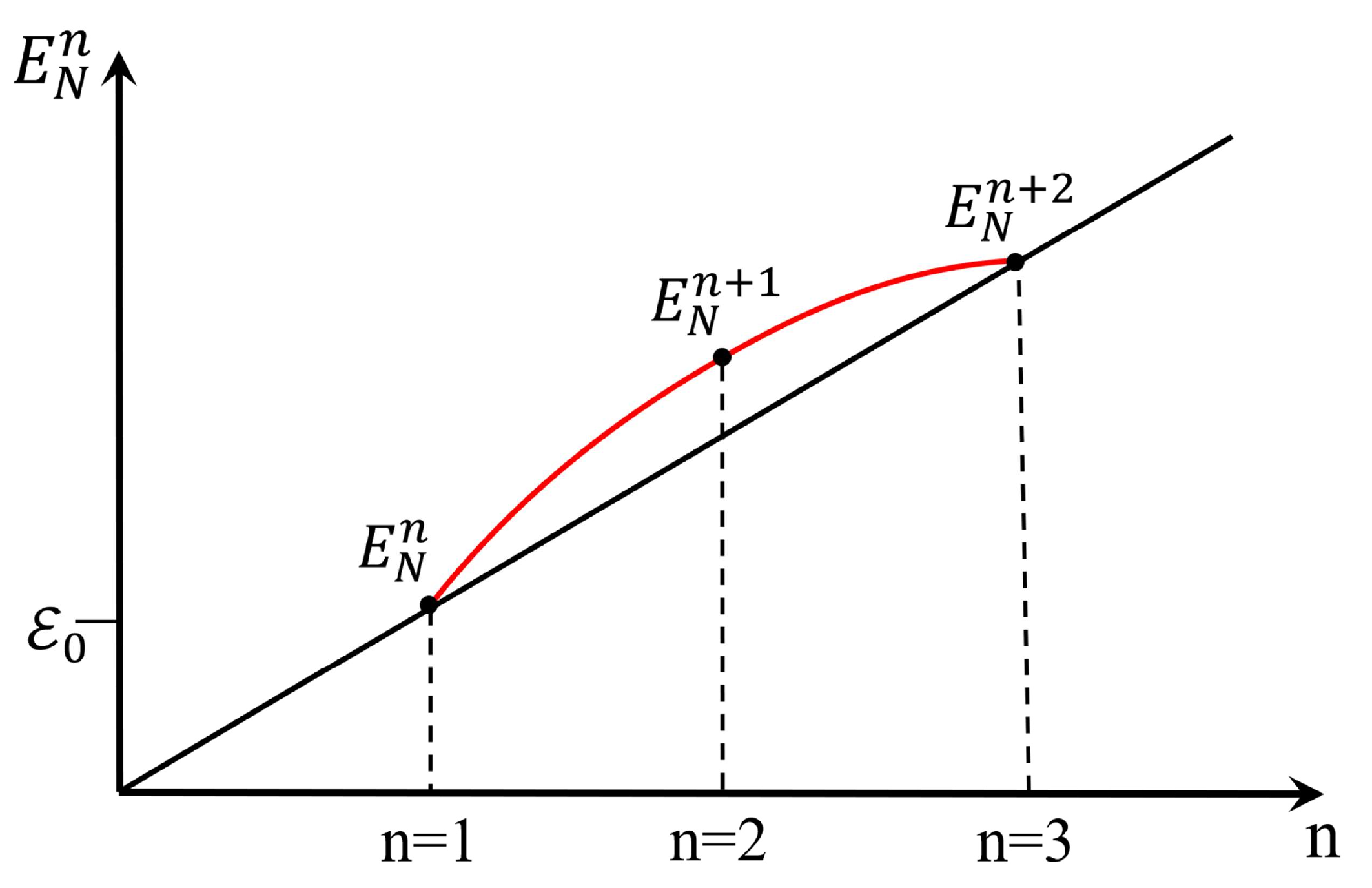}
\caption{A criterion, $E_N^n+E_N^{n+2} <2E_N^{n+1}$,  of the attraction of two electrons
 implies a concavity of the curve $E_N^n$. As a result, the population evolves as $n\rightarrow (n+2)$. }
\end{figure}
\label{F1}
\begin{figure}
\includegraphics[width = 90mm]{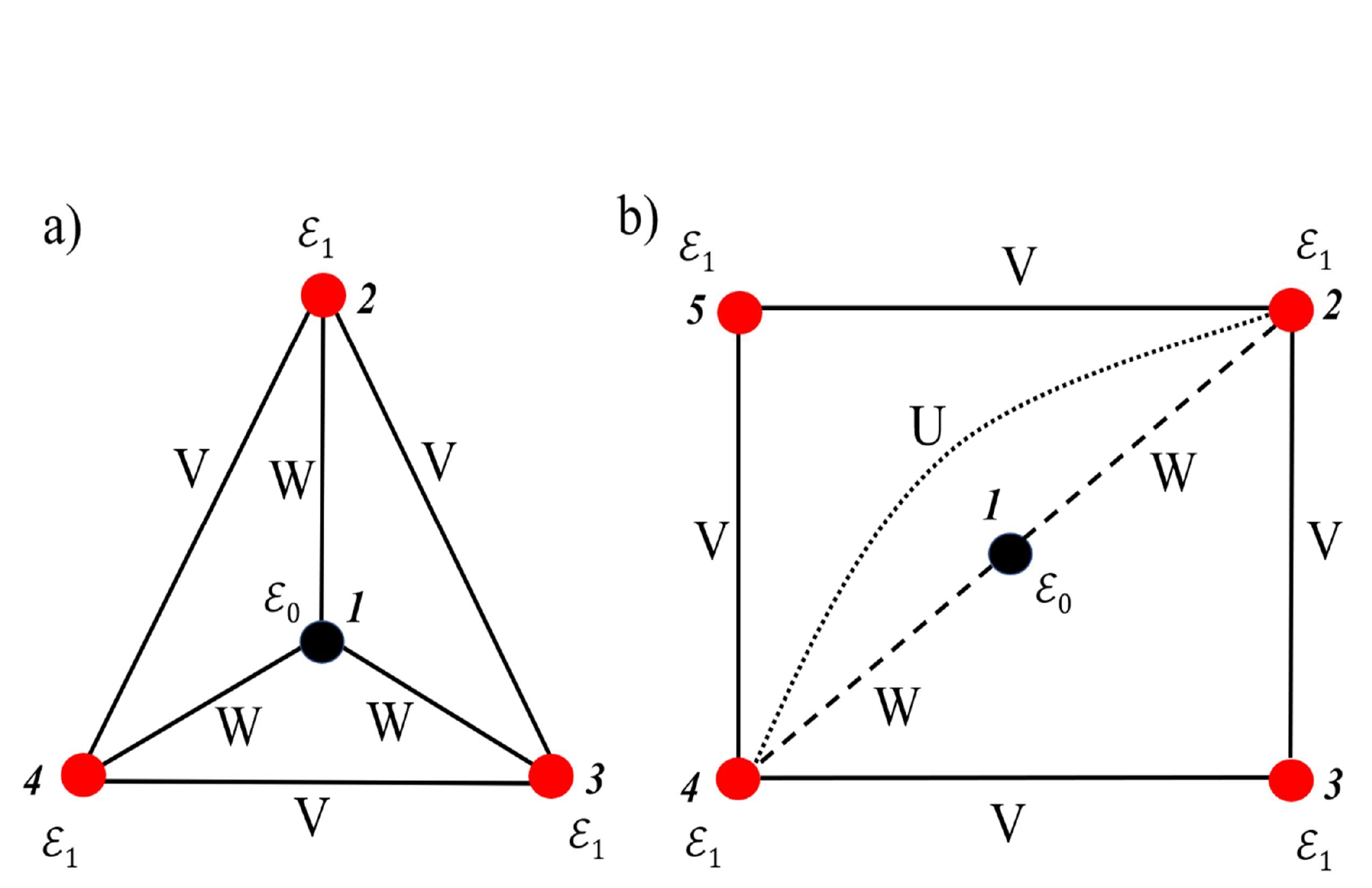}
\caption{a)  Illustration of the 4-site model. The energy of the central site, $\varepsilon_0$, is smaller than the energies of the corner sites, $\varepsilon_1$.
For a purely Coulomb system, the repulsions, $W$ and $V$,  are related as $3^{1/2}$. For a gate-screened interaction they are related as $3^{3/2}$; b) Illustration of the
5-site model.  The repulsions $W$, $U$, and $V$ are related as $2:2^{1/2}:1$ for a purely Coulomb interaction and as $2^{3}: 2^{3/2}:1$ for the gate-screened interaction.}
\label{F2}
\end{figure}
%Varma, Hamo, Ashoori+Heiblum, , Liang Fu

On the theory side, electron attraction in the repulsive systems
was invoked  to (i)  explain the  valence-skipping phenomenon, see e.g. Refs. \onlinecite{Varma,Dzero,Harrison,skipping,skipping1}, when the valence
of certain elements does not occur in the compounds which they form; (ii)
derive pairing within certain limits of the Hubbard model,\cite{Raghu}
which, essentially, amounts to replacing phonons by magnons.\cite{Batista}
Also, as demonstrated in Ref.~\onlinecite{LiangFu}, formation of charge-2e excitations, ``trimers", is favored in doped transition-metal dichalcogenide bilayers near the half-filling. The origin of pairing in Ref.  \onlinecite{LiangFu}
with two type of sites having different energy
is reduction of {\em electrostatic} energy which dominates over the kinetic energy.
Minimal model of a trimer requires four sites.
Note that such a 4-site model of pairing was previously studied in Ref. \onlinecite{Zhukov} in different relation.

\vspace{2mm}

On the experimental side, attention to the issue of  possible attraction of localized electrons was drawn by the
early experiments.\cite{Ashoori1992,Ashoori1993,Ashoori1997}
In these experiments, one-by-one  magneto-tunneling events of electrons from an electrode  into a big semiconductor island.
Certain portion of events revealed bunching of electrons into pairs.
Accounting this bunching by  attraction  of electrons  within a 4-site model
encounters a problem that incoherent  tunneling of two electrons takes too much time.
Later experiments\cite{Heiblum,Ashoori} suggested that two-electron events take place at the edge and are
related to the formation of  the edge states in   magnetic field.

Demonstration of pairing due to repulsion on a truly microscopic level was reported in Ref. \onlinecite{Hamo}.
Experimental setup in Ref. \onlinecite{Hamo} was very similar to the 4-site cluster (2-site polarizer and
 double-well quantum dot). In a  certain domain of gate voltages, one dot of a double-well system
was either empty or accommodated  a pair of electrons as a result of reoccupation of the dots
constituting the polarizer.

Due to flexibility of their  nanotube-based technique,
the authors of Ref. \onlinecite{Hamo} suggested several variants
for scaling-up their setup.
In particular, their approach can tackle the question
whether more complex many-particle processes
can be realized experimentally. In this paper we demonstrate
that, adding one extra site to the 4-site model, opens
a possibility of {\em three-electron} bunches.
%feasible
To establish a criterion for such three-electron bunches, we introduce
$E_N^n$, which is a minimal energy of $n$ electrons in a cluster of $N$ sites.
Next electron enters the cluster when the Fermi level position in surrounding system is equal to
\begin{equation}
\label{criterion1}
\mu_1=E_N^{n+1}-E_N^{n}.
\end{equation}
Two electrons enter the cluster at
\begin{equation}
\label{criterion2}
\mu_2=\frac{E_N^{n+2}-E_N^{n}}{2}.
\end{equation}
Formation of a $2e$ pair takes place when $\mu_2<\mu_1$, i.e. when the usual
condition $\big(E_N^{n+2}+E_N^n\big)<2E_N^{n+1}$, illustrated in Fig. \ref{F1}, is met.
Continuing the reasoning, three electrons enter the cluster at
\begin{equation}
\label{criterion3}
\mu_3=\frac{E_N^{n+3}-E_N^{n}}{3}.
\end{equation}
Then a $3e$-bunch is favored over single-electron and $2e$-bunch
under the conditions $\mu_3<\mu_1$ and $\mu_3<\mu_2$.

%involving sequential tunneling and rearrangements of the
%electronic system in the bulk [39,40] or at the edge [41] of a
%2D system, place two peaks at the same gate voltage.
%However, such models would result in diminished tunneling rates in our experiment. To have two electrons tunneling back and forth between the dot and the tunneling
%electrode at the exact same gate voltage, the first electron in
%a pair must first tunnel into the dot with less than the
%required energy
%ΔE
%to produce the rearrangement.
%Therefore, the second electron in the pair must tunnel into
%the dot at a rate that is fast compared to
%ΔE=ℏ (∼1011 s−1
%for the
%≈0.4 meV
% energy barrier as seen from the gate
%voltage spacing of single-electron peaks). As tunneling
%rates in our dots from the tunneling electrode are on the
%order of
%106 s−1,
%such negative-U models cannot explain
%our data. Another possibility would then be an effective
%zero repulsion between electrons in a pair, but we know of
%no model for this. The answer may lie in coherent tunneling
%of the two electrons. }

%%%%%%%%%%%%%%%%%%%%%%%%
\section{Two-electron bunching in a 4-site cluster}
For didactic reasons, in this section  we  review the steps unveiling  the two-electron  bunching  in the 4-site model.
 Extension to the 5-site model involves all the similar steps. On the other hand, the flow of logics offers an insight about possible extensions.

\noindent {\em Step I.}  \hspace{1mm}
A 4-site cluster is illustrated in Fig. \ref{F2}a.
We examine the evolution of the minimal energy of configurations with varying $n$.
The 4-site cluster is completely filled when n=4, i.e. only one configuration is possible. It is also obvious that,  for $n=1$,
the minimal energy corresponds to filling the central site with energy $\varepsilon_0<\varepsilon_1$.  Thus  $E_4^1=\varepsilon_{0}$.
When $n=2$,  the electron in the center can either stay there or move to the vertex in order to reduce the potential energy. Thus,  the candidates for $E_4^2$ are
\begin{equation}
\label{n=2}
 {\tilde  E}_{4}^2=\varepsilon_0+\varepsilon_1+W ~ \text {or}~  {\tilde{\tilde E}}_4^2= 2\varepsilon_{1}+V.
\end{equation}
Equally, there are two candidates to minimize the energy when the population of the cluster is $n=3$, namely
\begin{equation}
\label{n=3}
 {\tilde  E}_{4}^3= \varepsilon_0+2\varepsilon_1+2W+V  ~ \text {or}~  {\tilde{\tilde E}}_4^3=3\varepsilon_{1}+3V.
\end{equation}

\noindent{\em Step II.}
At this step, we make two  crucial assumptions: $ {\tilde  E}_{4}^2< {\tilde{\tilde E}}_4^2$ and
$ {\tilde  E}_{4}^3 >{\tilde{\tilde E}}_4^3$, which translate into the following inequalities
\begin{align}
\label{favorable}
&2\varepsilon_1+V> \varepsilon_0+\varepsilon_1+W,\nonumber\\
&3\varepsilon_1+3V<\varepsilon_0+2\varepsilon_1+V+2W.
\end{align}
These inequalities imply that  it is energetically unfavorable for two electrons to vacate the center, but it is favorable for three electrons to vacate the center.
Note that Eq. (\ref{favorable})  can be presented in a concise form
\begin{equation}
\label{assumptions}
1-\frac{V}{W}<\frac{\varepsilon_1-\varepsilon_0}{W}< 2\bigg(1-\frac{V}{W}\bigg).
\end{equation}
When the conditions Eq. (\ref{favorable}) are met, we have
\begin{equation}
\label{TwoEnergies}
 E_4^2=\tilde{E_4^2}, ~  E_4^3={\tilde{\tilde E}}_4^3.
\end{equation}
\noindent{\em Step III.}
At this step we require that the dependence $E_4^n$ is  ``concave",  as illustrated in Fig. \ref{F1}.
Using Eq. (\ref{TwoEnergies}), the pairing condition $E_4^1+E_4^3<2E_4^2$
takes the form

\begin{equation}
\label{checking}
\varepsilon_0 + 3\big(\varepsilon_1+V\big)<2\big(\varepsilon_0+\varepsilon_1+W\big).
\end{equation}
Note that the above condition
restricts the energy difference between the
center and the vertex sites: $\frac{\varepsilon_1-\varepsilon_0}{W}<2-\frac{3V}{W}$.

\begin{figure}
\includegraphics[width=90mm]{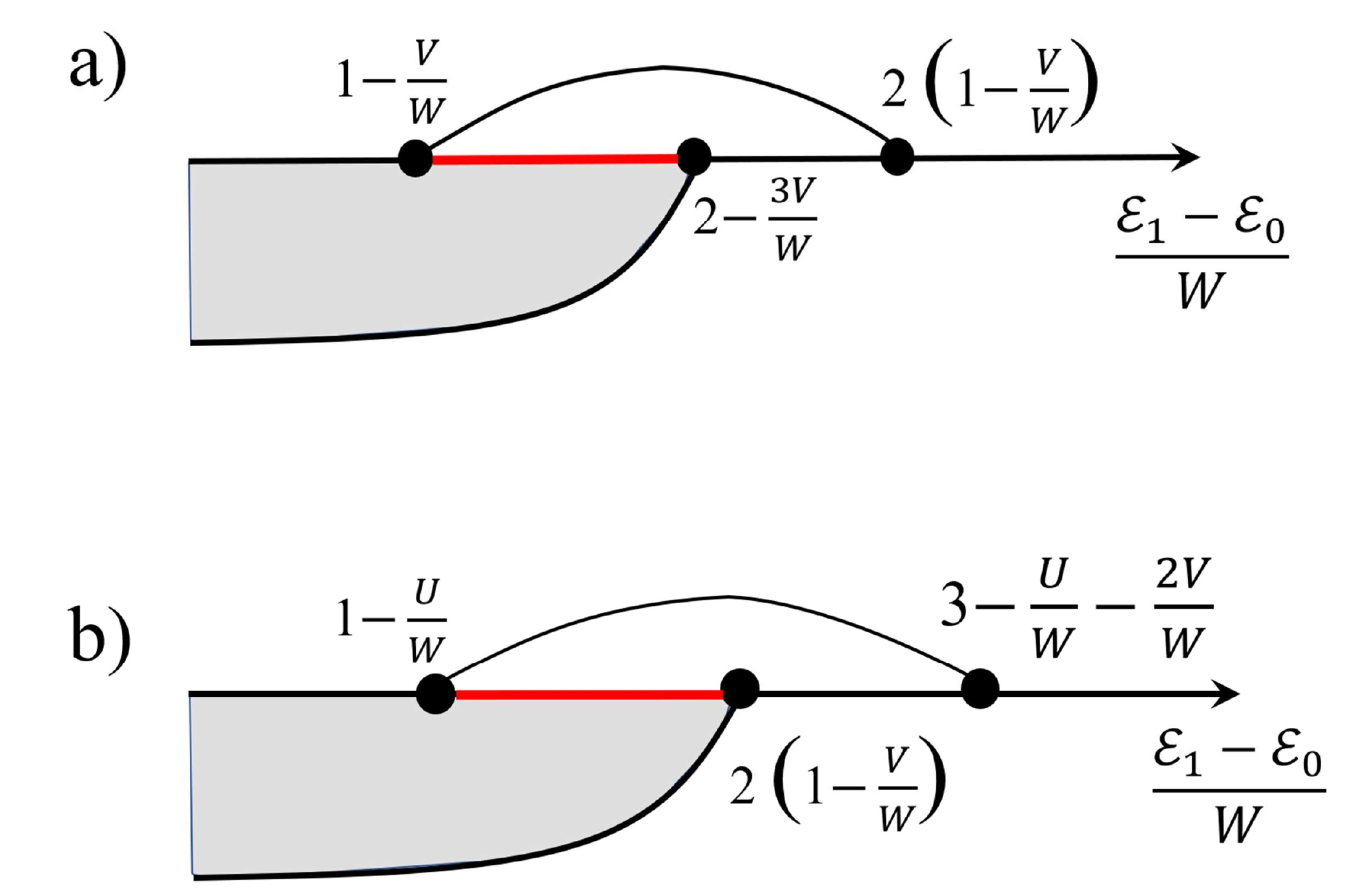}
\caption{Different domains of the dimensionless
imbalance, $\frac{\varepsilon_1-\varepsilon_0}{W}$, of the
site energies: a) 4-site model. In the domain of imbalances,
$\Big\{1-\frac{V}{W}, 2\big(1-\frac{V}{W}\big)\Big\}$ the energy
of two-electron configuration is minimal when one electron
occupies the central site, while for three-electron
configuration the energy is minimal when all three electrons
reside in the vertices. In the grey region the
condition of concavity, illustrated in Fig. \ref{F1}, is satisfied, so
that 2e-pairing occurs in the domain marked with red.
b) 5-site model. In the red domain the occupation
of the cluster with chemical potential evolves as
$n=1\rightarrow n=4$. }
\label{F4}
\end{figure}

\noindent{\em Step IV.}  We now go back to the assumptions made above,  Eq. (\ref{assumptions}),
and test whether they are consistent with concavity. This test is illustrated in Fig. \ref{F4}.
We see that the second inequality is satisfied automatically. To satisfy the fist inequality,
one needs $V<\frac{W}{2}$. As seen from Fig. \ref{F2}a, for purely Coulomb interaction,
the relation between $V$ and $W$ is $V=\frac{W}{3^{1/2}}$. Thus,
the first condition is not satisfied. It can be, however, satisfied if the gate
is present at a distance, $d$, above the plane of the cluster. Then the Coulomb interaction,
$V(\rho)=\frac{e^2}{\rho}$, gets modified to
$V(\rho)={e^2}\Big[\frac{1}{\rho}-\frac{1}{\left(\rho^2+4d^2\right)^{1/2}}\Big]$.
Then, for $\rho \gg d$, we have $V=\frac{W}{3^{3/2}}< \frac{W}{2}$, so that in the domain
\begin{equation}
\label{domain}
W-V<\varepsilon_1-\varepsilon_0<2W-3V,
\end{equation}
shown in Fig. \ref{F4}a with red, 2e-pairing is possible.

\begin{figure}
\includegraphics[width=90mm]{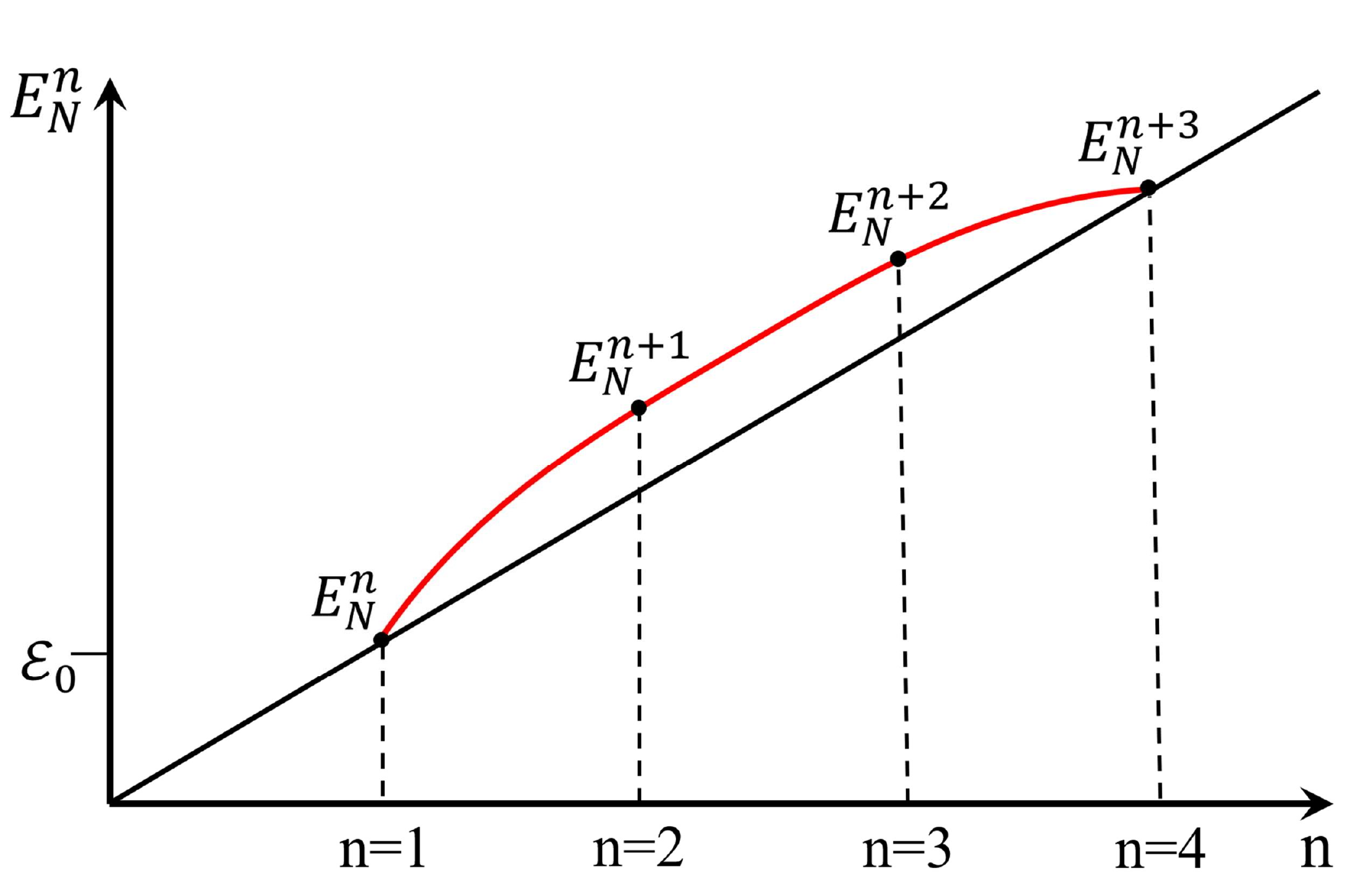}
\caption{Illustration of three-electron bunching in the cluster
of $N$ sites. The bunch
is possible under the conditions
$\frac{1}{3}\left(E_N^{n+3}-E_N^n\right)
<\frac{1}{2}\left(E_N^{n+2}-E_N^{n}\right)<E_N^{n+1}-E_N^n$.}
\label{F3}
\end{figure}

\section{Three-electron bunching in a 5-site cluster}

We now generalize the reasoning from the previous section to the
5-site model illustrated in Fig. \ref{F2}b. The central site is
surrounded by four vertexes. Three repulsion energies, shown in the
figure,  are related as $U<V<W$ in accordance with distances between the
corresponding sites. Similarly to the 4-site model, we assume that the
energy, $\varepsilon_0$, of the central site is lower than $\varepsilon_1$-the energy of the vertex sites. Obviously, $E_5^1=\varepsilon_0$. Our strategy in a search for three-electron
bunching  is based on the argument that, by placing electrons
in the vertexes, the configuration loses in its on-site energy, but
gains due to the reduction in potential energy.

\noindent {\em Step I.}  \hspace{1mm}
There are three candidates for the double occupation of the cluster. They are
\begin{equation}
\label{n1=2}
 {\tilde  E}_{5}^2=\varepsilon_0+\varepsilon_1+W ~ \text {or}~  {\tilde{\tilde E}}_5^2= 2\varepsilon_{1}+V
~\text {or}~
\Big( {\tilde{\tilde{\tilde E}}}\Big)_5^2=2\varepsilon_{1}+U.
\end{equation}
Since $V>U$, the energy ${\tilde  E}_{5}^2$ is smaller than $ {\tilde{\tilde E}}_5^2$ leaving us with only two candidates.
One can also see that there are three candidates for the triple occupation, namely
\begin{align}
\label{n1=3}
 &{\tilde  E}_{5}^3=3\varepsilon_1+2V+U ~ \text {or}~  {\tilde{\tilde E}}_5^3= 2\varepsilon_{1}+\varepsilon_0+2W+U,
~\text {or}~\nonumber\\
&\Big( {\tilde{\tilde{\tilde E}}}\Big)_5^3=2\varepsilon_{1}+\varepsilon_0+2W+V.
\end{align}
We can now compare ${\tilde{\tilde E}}_5^3$ to $\Big( {\tilde{\tilde{\tilde E}}}\Big)_5^3$ and realize that the latter has higher
energy since $V>U$. This, again, leaves us with two candidates for ${\tilde  E}_{5}^3$.

Finally, the occupation of the cluster with four electrons is possible in  two configuations  having different energies. These energies  are
\begin{equation}
\label{n1=4}
  {\tilde E}_5^4=\varepsilon_0+3\varepsilon_1+3W+2V+U, ~ \text{or} ~ {\tilde {\tilde E}}_5^4=4\varepsilon_1+4V+2U.
\end{equation}
Obviously,  the state with energy,  ${\tilde E}_5^4$ is four-fold degenerate depending on which vertex is empty.

\noindent {\em Step II.}  \hspace{1mm}
At this point, we make three crucial assumptions:

\begin{enumerate}
\centering
\label{inequalities}
    \item ${\tilde{\tilde E}}_5^2>{\tilde E}_5^2,$
    \item $ {\tilde{\tilde E}}_5^3>{\tilde E}_5^3,$
    \item ${\tilde {\tilde E}}_5^4<{\tilde E}_5^4.$
\end{enumerate}
The assumptions are made in order to ensure that the ground states with $n=2$ and $n=3$ include the center site, while
the ground state with $n=4$ includes the sites in the vertexes. Using Eqs. (\ref{n1=2}),  (\ref{n1=3}),   (\ref{n1=3}), the above
assumptions can be rewritten as

\begin{align}
\label{Favorable}
&\varepsilon_0+\varepsilon_1+W<2\varepsilon_{1}+U \Rightarrow  1-\frac{U}{W}< \frac{\varepsilon_1-\varepsilon_0}{W},\\
& \nonumber\\
&\varepsilon_0+2\varepsilon_1+2W+U>3\varepsilon_1+2V+U \nonumber\\
&\Rightarrow   2\Big( 1-\frac{V}{W}  \Big)  >   \frac{\varepsilon_1-\varepsilon_0}{W}, \\
&\nonumber\\
&\varepsilon_0+3\varepsilon_1+3W+2V+U>4\varepsilon_1+4V+2U\nonumber\\
&\Rightarrow  3-\frac{2V}{W}-\frac{U}{W}>      \frac{\varepsilon_1-\varepsilon_0}{W}     .
\end{align}
%These inequalities again imply that it is energetically unfavorable for two electrons to vacate
%the center, but we also now see that it is energetically favorable for both 3 and 4 electrons to vacate
%the center.
Now it is convenient to restructure the above inequalities into two separate ranges as

\begin{align}
\label{ranges}
&1-\frac{U}{W}<\frac{\varepsilon_1-\varepsilon_0}{W}<2\bigg(1-\frac{V}{W}\bigg),\nonumber\\
&1-\frac{U}{W}<\frac{\varepsilon_1-\varepsilon_0}{W}<3-2\frac{V}{W}-\frac{U}{W}.
\end{align}

Whether the three-electron bunches are allowed or not depends on whether or not the domains Eq. (\ref{ranges}) overlap, as illustrated in Fig. \ref{F4}b.
After making the above assumptions we can specify the ground-state configuration for each $n$, namely

%When the conditions (\ref{Favorable}) are met, we obtain
\begin{align}
\label{Threeenergies}
    E_5^2={\tilde E}_5^2, ~ E_5^3={\tilde E}_5^3, ~ E_5^4={\tilde{\tilde E}}_5^4
\end{align}

\noindent {\em Step III.}  \hspace{1mm}
We now require that the ground-state energies $E_5^n$ are arranged as shown in Fig. \ref{F3},
or, in other words, we require that these energies satisfy the conditions
\begin{equation}
\label{Domain}
\frac{1}{3}\left(E_5^{4}-E_5^1\right)
<\frac{1}{2}\left(E_5^{3}-E_5^{1}\right)<E_5^{2}-E_5^1.
\end{equation}
The first and the second  conditions  can be cast into a traditional form
\begin{align}
\label{Traditional}
&\frac{\varepsilon_1-\varepsilon_0}{W}>2\frac{V}{W}+\frac{U}{W},\nonumber\\
&\frac{\varepsilon_1-\varepsilon_0}{W}<3-4\frac{V}{W}-2\frac{U}{W}.
\end{align}
We see that, in the same way as Eq. (\ref{ranges}), the necessary requirements for 3-e bunches restrict the
asymmetry in single-electron energies both from below and from the above.
%\section{Acknowledgements}

\noindent {\em   IV.}  \hspace{1mm}
Now a nontrivial question arises: is there a domain in which the conditions Eq. (\ref{ranges}) and Eq. (\ref{Traditional}) are consistent with
each other? It  is apparent that for purely Coulomb repulsion the requirements Eq. (\ref{Traditional}) cannot be met. 
Indeed, with
purely Coulomb interaction, one has $\frac{U}{W}=\frac{1}{2}$, while  $\frac{V}{W}=\frac{1}{2^{1/2}}$. Then the right-hand side in the second inequality Eq. (\ref{Traditional}) is {\em negative}.
%{\bf refer to Fig. \ref{F3}b}
Turning to screened Coulomb repulsion, we have $\frac{U}{W}=\frac{1}{8}$,
while $\frac{V}{W}=\frac{1}{2^{3/2}}$. Then Eq. (\ref{ranges}) places the
asymmetry, $\frac{\varepsilon_1-\varepsilon_0}{W}$, into the interval
$\big(0.875, 1.29\big)$. At the same time, Eq. (\ref{Traditional}) restricts
the asymmetry to the interval $\big(0.83, 0.95\big)$. We see that two restricting
intervals overlap proving that 3e-bunches are allowed.

\section{Conclusion}
Certainly the conditions for the formation of 3e-bound state in the repulsive system are more restrictive than the
conditions for the formation of the 2e-boumd state.
We do not know whether the bunches containing more than three electrons are possible, but our finding motivates to search for them.

\end{document}